\documentclass[twocolumn,english,aps,prl,showpacs,superscriptaddress,groupaddress,floatfix,graphics,graphicx]{revtex4-1}
\usepackage{lmodern}
\usepackage{lmodern}
\usepackage[T1]{fontenc}
\usepackage[latin9]{inputenc}
\usepackage{geometry}
\usepackage{color}
\usepackage{babel}
\usepackage{float}
\usepackage{bm}
\usepackage{amsmath}
\usepackage{amssymb}
\usepackage{graphicx}
\usepackage[unicode=true,pdfusetitle,
 bookmarks=true,bookmarksnumbered=false,bookmarksopen=false,
 breaklinks=false,pdfborder={0 0 1},backref=false,colorlinks=false]
 {hyperref}
\hypersetup{
 colorlinks,linkcolor=blue,citecolor=blue,urlcolor=blue}

\makeatletter

\providecommand{\tabularnewline}{\\}

\usepackage{babel}

\usepackage{blindtext}
\usepackage{dsfont}\usepackage{bbm}\IfFileExists{lmodern.sty}{\usepackage{lmodern}}{}
\usepackage{babel}

\makeatother

\begin{document}

\title{Impact of complex adatom\textendash induced interactions on quantum
spin Hall phases }

\author{Flaviano~José~dos~Santos}

\affiliation{Peter Grünberg Institut and Institute for Advanced Simulation, Forschungszentrum
Jülich \& JARA, D-52425 Jülich, Germany}

\affiliation{RWTH Aachen University, D-52056 Aachen, Germany}

\author{Dario A. Bahamon}

\affiliation{MackGraphe \textendash{} Graphene and Nano-Materials Research Center,
Mackenzie Presbyterian University, Rua da Consolação 896, 01302-907,
São Paulo, SP, Brazil}

\author{Roberto B. Muniz }

\affiliation{Instituto de Física, Universidade Federal Fluminense, Niterói, Brazil}

\author{Keith~McKenna}

\affiliation{Department of Physics, University of York, York YO10 5DD, United
Kingdom}

\author{Eduardo~V. Castro}

\affiliation{CeFEMA, Instituto Superior Técnico, Universidade de Lisboa, Av. Rovisco
Pais, 1049-001 Lisboa, Portugal}

\affiliation{Beijing Computational Science Research Center, Beijing 100084, China}

\author{Johannes~Lischner}
\email{jlischner597@gmail.com}

\affiliation{Department of Physics and Department of Materials, and the Thomas
Young Centre for Theory and Simulation of Materials, Imperial College
London, London SW7 2AZ, United Kingdom}

\author{Aires~Ferreira}
\email{aires.ferreira@york.ac.uk}

\affiliation{Department of Physics, University of York, York YO10 5DD, United
Kingdom}
\begin{abstract}
\textcolor{black}{Adsorbate engineering offers a seemingly simple
approach to tailor spin\textendash orbit interactions in atomically
thin materials and thus to unlock the much sought-after topological
insulating phases in two dimensions. However, the observation of an
Anderson topological transition induced by heavy adatoms has proved
extremely challenging despite substantial experimental efforts. Here,
we present a multi-scale approach combining advanced first-principles
methods and accurate single-electron descriptions of adatom\textendash host
interactions using graphene as a prototypical system. Our study reveals
a surprisingly complex structure in the interactions mediated by random
adatoms, including hitherto neglected hopping processes leading to
strong valley mixing. We argue that the unexpected intervalley scattering
strongly impacts the ground state at low adatom coverage, which would
provide a compelling explanation for the absence of a topological
gap in recent experimental reports on graphene. Our conjecture is
confirmed by real-space Chern number calculations and large-scale
quantum transport simulations in disordered samples. This resolves
an important controversy and suggests that a detectable topological
gap can be achieved by increasing the spatial range of the induced
spin\textendash orbit interactions on graphene, e.g., using nanoparticles.}
\end{abstract}
\maketitle
The attachment of adsorbates to two-dimensional  materials has attracted
much interest in recent years, as a route to tailoring material properties
and realising novel phenomena \cite{Adorb_G_abinitio_Peeters_08,PhotoLum_TMD_engineering,Adsorb_G_Review_Lee_12,Adsorb_Various_TMD_Voiry_15,Adsorb_2D_Review_Ryder_16}.
In graphene, adatoms have been shown to induce band gaps \cite{BGap1_Fluorographene_Zhu_10,BGap2_Fluorographene_Nair_10,BGap3_Fluorographene_Snow_10,BGap4_Fluoride_DFT_Peeters_10,BGap5_Aryl_Lau_11},
magnetic moments \cite{G_MAGN_Ochoa_08,G_MAGN_Nair_12,G_MAGN_Herrero_16}
and even superconductivity \cite{Supercond_G_Profefa_12,Supercond_G_Li_13,Supercond_G_Ludbrook,Supercond_G_Chapman_16}. 

\textcolor{black}{Adsorbate engineering could likewise provide atomic
control over fundamental spin\textendash orbit phenomena, such as
 spin relaxation \cite{SpinRelax_G_Hernando_09,SpinRelax_G_Pi_10,SpinRelax_G_Federov_13,SpinRelax_G_Tuan_14}
and Mott (skew) scattering \cite{SHE_G_Pachoud_14,SHE_Graphene_Ferreira_14,SHE_Graphene_Ozyilmaz_Ferreira_14,SHE_Huang_16}.
}Recent studies have predicted that the dilute assembly of heavy adatoms
can massively enhance the weak spin\textendash orbit energy gap of
graphene \cite{QAHE_G_Adatom_Zhang,QSH_G_Adatom_Weeks_11,QSH_G_Adatom_Jiang_12,QSH_G_Adatom_Brey},
opening a promising route towards the realisation of nontrivial topological
insulating phases, including the quantum spin Hall (QSH) state \cite{QSH_G_Kane_Mele_05}.
However, transport measurements on samples decorated with heavy species,
including In and Ir, have yet to show any signature of topological
gap opening \cite{QSH_Exp_Adatom_1,QSH_Exp_Adatom_2,QSH_Exp_Adatom_3,QSH_Exp_Adatom_4,QSH_Exp_Adatom_5}.\textcolor{black}{{}
In this work, we show that a thorough treatment of disorder\textemdash combining
accurate model Hamiltonians with quantum transport simulations\textemdash is
essential to predict the topological character of adatom-engineered
systems and reconcile this contradiction. Our approach reveals that
randomly distributed heavy adatoms on graphene give rise to }scattering
between inequivalent valleys in the band structure, hindering the
emergence of topologically protected edge states \emph{even in the
absence of extrinsic factors}, such as adatom clustering \cite{QSH_clustering_2014}.
This resolves a controversy regarding the nature of spin\textendash orbit
interactions in adatom-decorated graphene and indicates that decoration
with small clusters or nanoparticles, for which intervalley scattering
is strongly reduced, may offer a route towards the realisation of
the much sought-after QSH phase. 

Pristine graphene is a QSH insulator, but the smallness of its intrinsic
spin\textendash orbit interaction (in the range of 25\textendash 50
$\mu$eV \cite{Intrinsic_SOC_G_Huertas-Hernando_06,Intrinsic_SOC_G_Min_06,Intrinsic_SOC_Konschuh_10,Intrinsic_SOC_Sichau_17})
has thusfar precluded achieving the dissipationless quantum transport
regime \cite{QSH_QWells_Konig_07,QSH_QWells_Inverted_Liu_08,QSH_QWells_Inverted_Du_15}.
The opening of a detectable QSH gap requires a massive enhancement
of graphene's characteristic spin\textendash orbit coupling (SOC),
which preserves spin angular momentum $S_{z}$ \cite{QSH_G_Kane_Mele_05}.
Previous work suggested that this can be achieved via decoration with
nonmagnetic adatoms with a $p$-outer electron shell \cite{QSH_G_Adatom_Weeks_11},
as they induce spin\textendash conserving 'intrinsic-like' SOC \cite{SHE_G_Pachoud_14}.
As a prototypical heavy element, we consider thallium (Tl) \cite{QSH_G_Adatom_Weeks_11}.
The quasiparticle band structure of the thallium adatom on graphene
was obtained employing a fully-relativistic \emph{ab initio }GW approach;
see Supplemental Material (SM) \cite{SM} for details. 

\emph{First-principles multi-scale approach.\textemdash }We use our
first-principles supercell calculations to parameterize a single-electron
Hamiltonian capturing \emph{all} the relevant interactions mediated
by (dilute) adatoms embedded in large area graphene samples. The first
step is to derive a graphene\textendash single-adatom tight-binding
(TB) model that faithfully reproduces the \emph{ab initio} band structure.
Enforcing time-reversal symmetry and invariance with respect to the
$C_{6v}$ point group, one easily finds $H=H_{g}+H_{a}+V_{ga}$ \cite{SHE_G_Pachoud_14,QSH_G_Adatom_Weeks_11},
where 

\begin{align}
H_{g} & =-t\sum_{\langle ij\rangle}c_{\bm{r}_{i}}^{\dagger}c_{\bm{r}_{j}}+(\delta H_{t',t''}-\delta\mu\sum_{i\in P}c_{\bm{r}_{i}}^{\dagger}c_{\bm{r}_{i}})\,,\label{eq:H_g}\\
H_{a} & =\sum_{m=0,\pm1}\epsilon_{|m|}d_{m}^{\dagger}d_{m}+\lambda(d_{1}^{\dagger}s^{z}d_{1}-d_{-1}^{\dagger}s^{z}d_{-1})\,+\nonumber \\
 & \qquad\qquad\,\,\sqrt{2}\lambda(d_{0}^{\dagger}s^{-}d_{-1}+d_{0}^{\dagger}s^{+}d_{1}+\textrm{H.c.})\,,\label{eq:H_a}\\
V_{ga} & =-\sum_{m=0,\pm1}(i^{|m|}\tau_{|m|}\Omega_{m}^{\dagger}d_{m}+\textrm{H.c.})\,.\label{eq:V}
\end{align}
The first two terms are the Hamiltonians of $\pi$-electrons on graphene
and 6$p$-states of a Tl atom, respectively. $c_{\bm{r}_{i}}^{\dagger}\equiv(c_{\bm{r}_{i}\,\uparrow}^{\dagger},c_{\bm{r}_{i}\,\downarrow}^{\dagger})$
and $d_{m}^{\dagger}\equiv(d_{m\,\uparrow}^{\dagger},d_{m\,\downarrow}^{\dagger})$
are th\textcolor{black}{e corresponding fermionic creation operators,
$s^{x,y,z}$ are Pauli matrices acting on the spin space and $s^{\pm}=(s^{x}\pm is^{y})/2$.
The sites adjacent to the adatom d}efine a hexagonal plaquette, \textcolor{black}{$P\equiv\{1,...,6\}$}.
$V_{ga}$ is the adatom\textendash graphene hybridization term written
as a function of the plaquette operator for states with definite angular
momentum, $\Omega_{m}^{\dagger}=(1/\sqrt{6})\sum_{j\in P}\exp{\{i\pi m(j-1)/3\}}c_{\bm{r}_{j}}^{\dagger}$
\cite{Note_Minimal_Model}. Next-nearest and third-nearest neighbor
corrections ($\delta H_{t',t''}$) are included in order to improve
agreement to the first-principles results. 
\begin{figure}
\includegraphics[width=0.75\columnwidth]{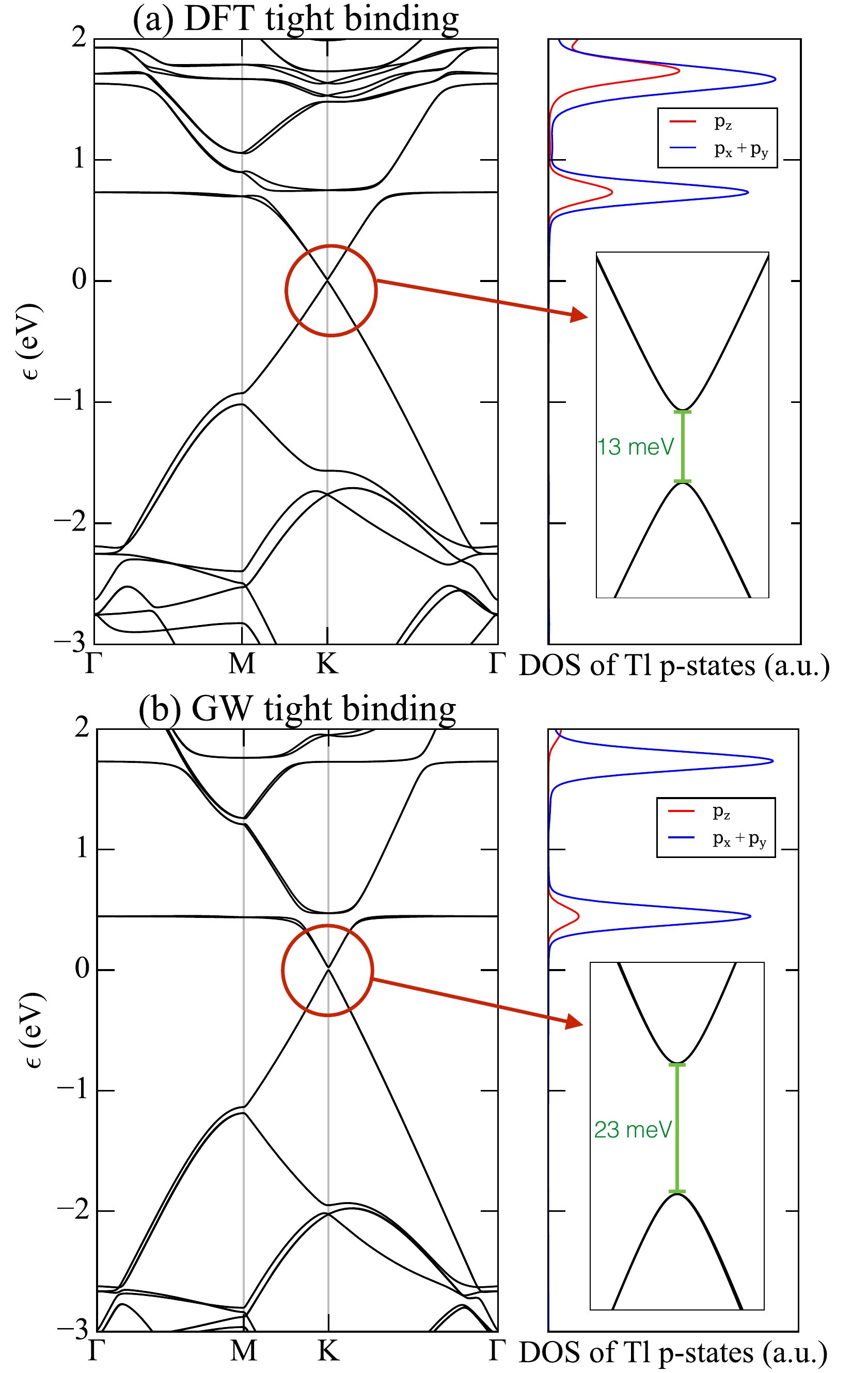} \caption{\label{fig:01} a) Fully-relativistic DFT band structure for a thallium
atom on graphene. b) The corresponding quasiparticle band structure
from a fully-relativistic GW calculation. }
\end{figure}
The minimal Hamiltonian {[}Eqs.~(\ref{eq:H_g})-(\ref{eq:V}){]}
contains 9 parameters: the C-C hoppings, $t$, $t'$ and $t''$, the
local chemical potential change on C sites next to Tl $\delta\mu$,
the Tl outer-shell energies, $\epsilon_{0}$ and $\epsilon_{\pm1}$,
the Tl spin-orbit energy $\lambda$ and the C-Tl hoppings, $\tau_{0}$
and $\tau_{\pm1}$. We adjust these parameters until the band structures
quantitatively reproduce the first-principles calculations over a
window of $\pm1$\,\,eV around the Dirac point (see SM \cite{SM}).
The initial guess for the parameters is informed by a direct evaluation
of hopping integrals between atom-centred maximally localized Wannier
functions \cite{SM}. The quasiparticle band structure obtained from
density functional theory (DFT)- and GW-parameterized minimal TB models
is shown in Fig.\,\ref{fig:01}. Bands below the Dirac point ($\epsilon\equiv0$)
derive mostly from graphene $\pi$ states. The flat band with energy
$\approx0.4$\,\,eV is a Tl 6$p$ state. Interestingly, the GW corrections
are seen to bring this band \emph{closer} to the Dirac point. Moreover,
the gap at the Dirac point is 23\,\,meV \emph{significantly larger}
than the DFT value of 13\,\,meV. To what extent the optimistic first-principles
estimates signal a measurable topological gap in real samples will
depend on a delicate competition between spin-conserving SOC and two
other interactions mediated by disorder, which we unveil in what follows.

\begin{figure}
\includegraphics[width=0.65\columnwidth]{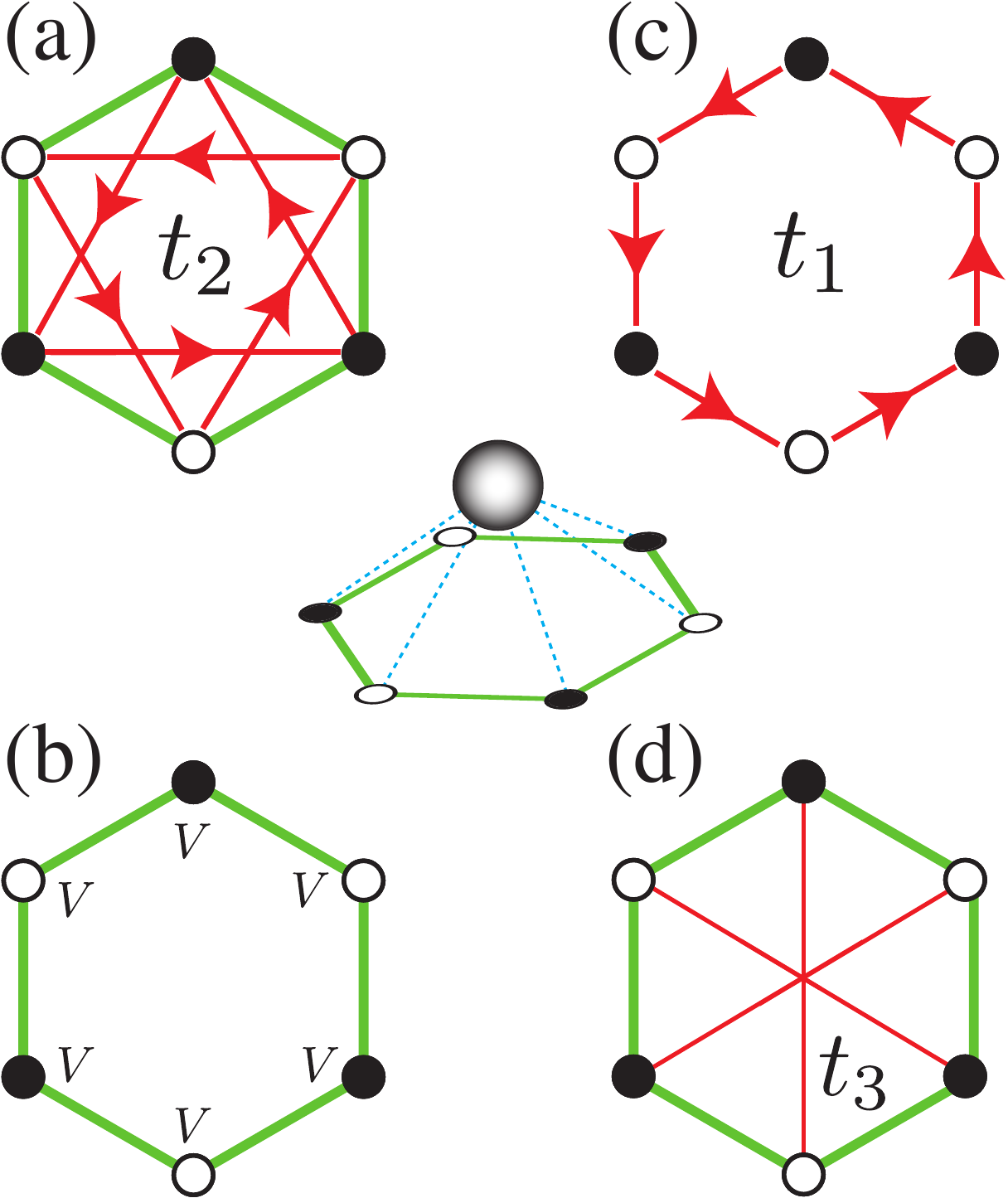} \caption{\label{fig:02} Complex adatom\textendash graphene interactions in
realistic scenarios. (a) NNN  hoppings. (b) On-site energies. (c-d)
Hopping processes opening the intervalley channel unveiled in this
work. Green and red lines represent bare and adatom-induced hoppings,
respectively; arrows indicate the presence of imaginary ``chiral''
components, making such hoppings sensitive to directions and spin
projections. }
\end{figure}

\emph{Adatom scattering potential.\textemdash }In realistic conditions,
dilute adatoms occupy random positions and thus act as \emph{scattering
centres}. The information on the adatom scattering potential is contained
in the local density of states (LDOS) \cite{LDOS_graphene_Bena_08}.
The crucial step in our multi-scale approach consists of deriving
a (graphene-\emph{only}) TB model for the scattering potential compatible
with the first-principles results for the adatom\textendash graphene
supercell (see Fig.~\ref{fig:01}). To this end, we trace out the
adatom degrees of freedom in Eqs.~(\ref{eq:H_g})-(\ref{eq:V}) through
a Löwdin transformation. Formally, the resulting graphene\textendash only
TB Hamiltonian is given by $H_{\textrm{eff}}=H_{g}+\Sigma_{P}(\epsilon)$,
where $\Sigma_{P}(\epsilon)$ is the real-space self-energy generated
by a single adatom \cite{SM}. This term breaks translational symmetry
and thus acts as a \emph{bona fide} disorder interaction. Finally,
the Hamiltonian of graphene with a dilute adatom coverage is obtained
by adding up independent contributions $\{\Sigma_{P}(\epsilon)\}$
from adatoms located at random plaquettes, $P_{k}$ ($k=1...N)$.
This procedure has two advantages. It captures \emph{all} short-range
interactions induced by the adatom (see below). Second, being based
on a graphene-only description, it allows a straightforward interpretation
of quantum transport calculations. Explicit evaluation of the self-energy
gives rise to the following effective interaction Hamiltonian, $\hat{V}_{\textrm{dis}}=\sum_{k=1}^{N}\hat{\Sigma}_{P_{k}}$,
with
\begin{align}
\hat{\Sigma}_{P} & =\sum_{\langle ij\rangle\in P}\,\,\left(t_{\text{1}}^{\prime}c_{\boldsymbol{r}_{i}}^{\dagger}c_{\boldsymbol{r}_{j}}+\imath\zeta_{ij}t_{1}^{\prime\prime}c_{\boldsymbol{r}_{i}}^{\dagger}s^{z}c_{\boldsymbol{r}_{j}}\right)+\nonumber \\
 & \quad\sum_{\langle\langle ij\rangle\rangle\in P}\left(t_{\text{2}}^{\prime}c_{\boldsymbol{r}_{i}}^{\dagger}c_{\boldsymbol{r}_{j}}+\imath\zeta_{ij}t_{2}^{\prime\prime}c_{\boldsymbol{r}_{i}}^{\dagger}s^{z}c_{\boldsymbol{r}_{j}}\right)+\nonumber \\
 & \;\,\;\sum_{\langle\langle\langle ij\rangle\rangle\rangle\in P}t_{\text{3}}^{\prime}c_{\boldsymbol{r}_{i}}^{\dagger}c_{\boldsymbol{r}_{j}}+\lambda_{R}\sum_{i,j\in P}c_{\bm{r}_{i}}^{\dagger}\Lambda_{ij}c_{\bm{r}_{j}}\,,\label{eq:Vimp}
\end{align}
where $\zeta_{ij}$ equals $\mp1$ for circulation around the $P$-th
plaquette (anti-)clockwise and $\Lambda_{ij}=\imath(\xi_{ij}s^{+}+\xi_{ij}^{*}s^{-})$
with $\xi_{ij}=\exp[\imath(j-1)\pi/3]-\exp[\imath(i-1)\pi/3]$. The
effective hoppings are defined by $\lambda_{R}=\sqrt{2}\lambda\tau_{0}\tau_{1}/D_{\textrm{ad}}$
and 
\begin{align}
t_{n}^{\prime}+\imath t_{n}^{\prime\prime} & =e^{\imath n\pi/3}X_{1}+X_{0}+e^{-\imath n\pi/3}X_{-1}\,,\label{eq:tn}
\end{align}
where $X_{1}=\tau_{1}^{2}(\epsilon-\epsilon_{0})/D_{\textrm{ad}}$,
$X_{0}=\tau_{0}^{2}(\epsilon-\epsilon_{1}+\lambda)/D_{\textrm{ad}}$,
$X_{-1}=\tau_{1}^{2}/[6(\epsilon-\epsilon_{1}-\lambda)]$ and $D_{\textrm{ad}}=6[\epsilon^{2}-\epsilon(\epsilon_{0}+\epsilon_{1}-\lambda)+\epsilon_{0}(\epsilon_{1}-\lambda)-2\lambda^{2}]$.
The first terms ($t_{1}^{\prime}$ and $t_{1}^{\prime\prime}$) modify
the hopping between nearest-neighbor (NN) atoms, while the second
line describes next-nearest-neighbor (NNN) hoppings, including a chiral
component ($t_{2}^{\prime\prime}$), which\textemdash in the \emph{absence}
of the other terms\textemdash transforms graphene into a QSH insulator
\cite{QSH_G_Adatom_Weeks_11}. The terms in the third line capture
hoppings between C atoms on opposite sides of the adatom ($t_{3}$)
and spin-flip processes between all pairs of sites in the impurity
plaquette ($\lambda_{R}$). The latter is a Rashba-type interaction,
which is vanishing small near the Dirac points and thus can be safely
neglected \cite{QSH_G_Adatom_Weeks_11,SHE_Graphene_Ferreira_14}.
The relevant interactions are visualized in Fig.~\ref{fig:02}. 

\begin{figure*}[t]
\centering \includegraphics[width=1\textwidth]{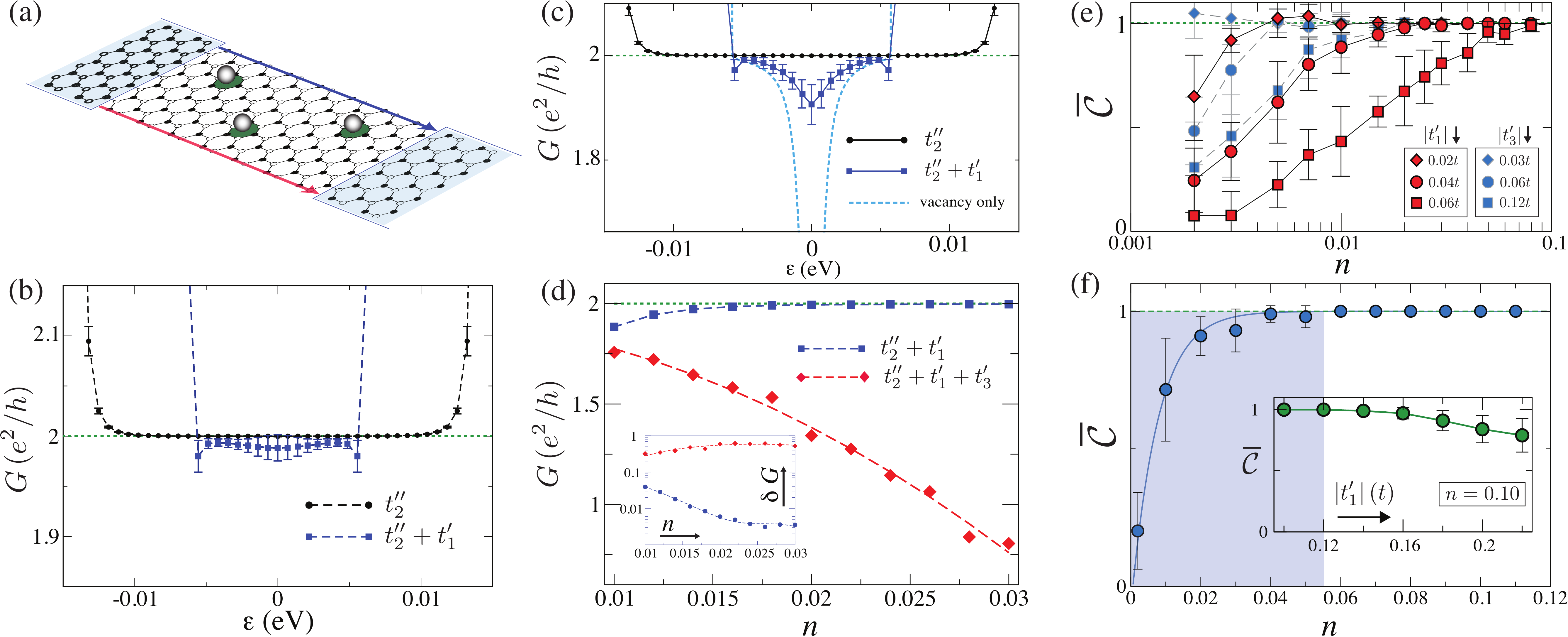}\caption{\label{fig:03} Topological properties of nanoribbon and bulk graphene
with random nonmagnetic heavy adatoms. \textcolor{black}{(a) Schematic
of a two-terminal device in the QSH regime; two edge states with opposite
spins contribute with $G=2e^{2}/h$. (b) Energy dependence of the
conductance for two distinct adatom\textendash graphene models (1\%
coverage). Parameters: $t=2.7$~eV, $t_{1}^{\prime}=-2.4$~eV and
$t_{2}^{\prime\prime}=-0.23$~eV. (c) Conductance curve when a single
vacancy is added to the adatom-decorated nanoribbon. Same parameters
as in (b). The conductance of the defected nanoribbon in the absence
of adatoms is also shown (dashed line). (d) Coverage dependence of
the conductance showing the degradation of the quantized plateau.
The behavior of fluctuations is shown in the inset. Parameters: same
as in (b) and (c) and $t_{3}^{\prime}=-1.0$~eV. (e) Disorder averaged
Chern number as function of adatom coverage for realistic spin\textendash orbit
coupling ($t_{2}^{\prime\prime}=0.01t$) and selected values of $t_{1}$
and $t_{3}$. (f) Coverage dependence of $\bar{\mathcal{C}}$ when
all the interactions are ``turned on'' using DFT base values: $t_{1}/t=-0.12-0.01\,i$,
$t_{2}^{\prime}/t=-0.07-i\,0.01$ and $t_{3}/t=-0.17$. Fitting to
$\bar{\mathcal{C}}(n)=\tanh(n/n^{*})$ yields a critical adatom coverage
$n^{*}\approx0.01$. The inset shows the dependence of $\bar{\mathcal{C}}$
with $t_{1}^{\prime}$ indicating a fast closing of the topological
gap upon increasing NN hopping correction. Twenty-four independent
adatom configurations were used in (b) and (c), and 100 in (d). For
simplicity, adatom-induced hoppings are fixed to their values at the
band center, $t_{n}(\epsilon=0)$, in all calculations.}}
\end{figure*}

Remarkably, the effective Hamiltonian obtained here by a rigorous
adatom-decimation procedure is \emph{far more complex than previous
models} \cite{QSH_G_Adatom_Weeks_11}. Importantly, $t_{1}^{\prime}$
and $t_{3}^{\prime}$ for heavy species can be significantly larger
than the chiral NNN hopping. From the \emph{ab initio} parameters
derived for Tl \cite{SM} we obtain $t_{1}^{\prime}\approx10t_{2}^{\prime\prime}$
at $\epsilon=0$. To shed further light on the significance of hitherto
neglected terms {[}(c)-(d) in Fig.~\ref{fig:02}{]}, we derive a
long-wavelength effective description. As customary, we introduce
the field operators, $c_{\sigma\tau s}(\boldsymbol{k})=\int d\boldsymbol{r}\,e^{\imath(\boldsymbol{k}+\tau K)\cdot\boldsymbol{r}}\,\Psi_{\sigma\tau s}(\boldsymbol{r})$,
\textcolor{black}{with $\sigma\,[\tau]=\pm1$ describing the projection
low-energy states} on the $A$($B$) sublattice {[}at $\boldsymbol{K}_{\pm}${]}
for \textcolor{black}{spin $s=\pm1$ }. Substituting in Eq.\,\,(\ref{eq:Vimp})
and performing a series expansion around the inequivalent Dirac points,
$\boldsymbol{K}_{\pm}=\pm K\hat{k}_{x}$, one obtains the effective
interaction: $\hat{V}_{P}=\Psi^{\dagger}(\boldsymbol{r})\hat{\mathcal{V}}_{P}(\boldsymbol{r})\Psi(\boldsymbol{r})$
\cite{Comment}, with 
\begin{eqnarray}
\hat{\mathcal{V}}_{P}(\boldsymbol{r})=(\Delta\tau_{z}\sigma_{z}s_{z}+g_{0}\tau_{x}\sigma_{x}+g_{1}\tau_{y}\sigma_{y}s_{z})f_{P}(\boldsymbol{r}),\label{eq:V_imp_continuum}
\end{eqnarray}
to leading order in $\boldsymbol{k}/K$, and where we omitted a scalar
term (see SM~\cite{SM}). $\boldsymbol{\sigma}$ and $\boldsymbol{\tau}$
denote Pauli matrices acting on sublattice and valley degrees of freedom,
respectively, and $f_{P}(\boldsymbol{r})\propto\delta(\boldsymbol{r}-\boldsymbol{r}_{P})$
describes the spatial profile of the adatom potential. The first term
derives from the NNN hopping, $\Delta=3\sqrt{3}t_{2}^{\prime\prime}$
\cite{QSH_G_Kane_Mele_05}. The remaining terms are scalar and spin\textendash orbit
interactions \emph{connecting valleys}, with strengths $g_{0}=3(t_{3}^{\prime}-t_{1}^{\prime})$
and $g_{1}=3\sqrt{3}t_{1}^{\prime\prime}$, respectively. For $p$-outer-shell
adatoms $t_{1}^{\prime\prime}=t_{2}^{\prime\prime}$ {[}see Eq.\,(\ref{eq:tn}){]}
and thus $g_{1}=\Delta$, i.e., \emph{intra- and inter-valley spin\textendash orbit
scattering processes must be considered on equal footing}. Based on
general arguments for disordered zero-gap semiconductors, one expects
that the mixing of states at inequivalent degeneracy points is detrimental
for the topological phase \cite{Hasan_Kane_10,Fradkin_86,Suzuura_Ando_02}.
\textcolor{black}{An estimation using DFT-optimized parameters gives
$g_{0}=0.41$\,eV and $g_{1}=\Delta=-0.11$\,eV at $\epsilon=0$.
}Such a dominance of intervalley processes in the coarse-grained description
is a strong indication that the topological gap displayed by Tl\textendash graphene
supercells (Fig.\,\ref{fig:01}) will be fragile in a disordered
scenario, which would naturally explain the negative experimental
results \cite{QSH_Exp_Adatom_1,QSH_Exp_Adatom_2,QSH_Exp_Adatom_3,QSH_Exp_Adatom_4,QSH_Exp_Adatom_5}.
This idea is reinforced by the fact that electrons in graphene are
very sensitive to valley mixing processes with an origin in short-range
impurities, as the respective Friedel oscillations are known to decay
as $1/r$ at large distances (as opposed to the $1/r^{2}$ law from
intravalley scattering \cite{LDOS_graphene_Bena_08}). 

\emph{Real-space quantum transport study.\textemdash }To investigate
the implications of the complex structure of the effective adatom
potential, we have carried out transport calculations within the Landauer-Büttiker
framework. In the QSH regime, a pair of counter-propagating gapless
edge states protected from elastic backscattering emerge at the interfaces
to vacuum {[}Fig.\,\,\ref{fig:03}\,(a){]} \cite{QSH_G_Kane_Mele_05}.
To probe the robustness of the extrinsic QSH insulating phase and
its concomitant helical edge states, we calculated the two-terminal
conductance of large armchair nanoribbons (width $W=313.6$~nm, and
length $L=298.2$~nm) with randomly distributed adatoms connected
to pristine graphene leads. The central channel contains in excess
of \emph{3.5 million atoms} and efficient recursion techniques are
employed to solve a system of this large size \cite{SM}. The smoking
gun for the topologically-protected edge states is the emergence of
quantized conductance $G=2e^{2}/h$ with a plateau width proportional
to the SOC strength \cite{supp15_Buttiker_98,Roth_09}.\textcolor{blue}{{}
}To probe the experimentally relevant adatom coverages would require
prohibitively large computational domains, in order to resolve the
typically small spin\textendash orbit gaps $\Delta E\approx0.3\,n$~(eV),
where $n$ is the adatom coverage. To overcome this difficulty, we
rescaled uniformly the effective hoppings, $t_{n}\rightarrow r\,t_{n}$
with $r=10$. The main findings are summarized in Fig.\,\,\ref{fig:03}.
When only the intrinsic-type SOC term is included ($t_{2}^{\prime\prime})$,
the two-terminal conductance exhibits a plateau at small energies
with $G=2e^{2}/h$ {[}Fig.\,\,\ref{fig:03} (b) (black dots){]}.
The variance $\Delta G$ is found to be \emph{zero} up to numerical
accuracy. Such a perfectly quantized effect shows that the nanoribbon
has been transformed into a QSH insulator. The plateau width precisely
saturates the upper bound $\Delta E_{\textrm{SOC}}\le2n|\Delta|$,
which is the topological gap obtained by Weeks \emph{et al} \cite{QSH_G_Adatom_Weeks_11}.
However, the plateau \emph{shrinks} when $t_{1}^{\prime}$ is turned
on (blue dots) indicating a closing of the topological gap. This effect
is accompanied by significative fluctuations, $\Delta G\approx\pm0.1\,e^{2}/h$,
showing that states delocalized through the nanoribbon contribute
now to the electronic transport. This\emph{ striking decay of helical
edge states into the bulk} is confirmed by the numerical evaluation
of the spin-polarized bond current density maps; see SM \cite{SM}.
The remaining adatom interactions affect the QSH phase in two different
ways: (i) \textcolor{black}{the imaginary NN hopping shifts the conductance
plateau (not shown) and (ii) the hopping $t_{3}^{\prime}$ enlarges
the central plateau and }\textcolor{black}{\emph{increases}}\textcolor{black}{{}
the fluctuations}. 

\textcolor{black}{We now provide compelling evidence that QSH phases
induced by random dilute adatoms are especially fragile in truly disordered
scenarios, where additional sources of scattering are unavoidable.
To this end, we introduce a topological defect (vacancy) in the nanoribbon,
by cutting all bonds adjacent to one carbon atom. }Vacancies introduce
(quasi)localized states at zero energy strongly impacting the graphene
electronic properties \textcolor{black}{\cite{ZEM_Pereira_06,ZEM_Ugeda_10,ZEM_Hafner_14,ZEM_Hafner_15,ZEM_Weik_16,ZEM_nanoribbon_15}}.
Given our choice of a metallic armchair nanoribbon, we locate the
vacancy on a site with finite density of states \textcolor{black}{\cite{GNR_defect_middle}}.
This choice guarantees that the vacancy works as a resonant scatter
(introducing mid-gap states), leading to a strong suppression of the
conductance at low energy, $G\rightarrow0$ as $\varepsilon\rightarrow0$
\textcolor{black}{{[}}Fig.~\ref{fig:03}(c)\textcolor{black}{{]}}.\textcolor{black}{{}
Adding a small coverage of idealized adatoms with only NNN hopping
($t_{2}^{\prime\prime}$) gives rise to the expected quantization
of the conductance, as the helical edge states resulting from the
SOC enhancement can perfectly avoid the vacancy (no backscattering).
Quite strikingly, when a NN hopping correction $t_{1}^{\prime}\approx10t_{2}^{\prime\prime}$
(typical of Tl adatoms in rows 5 and 6 of the periodic table) is turned
on, the conductance acquires its basic shape prior to adatom decoration,
unambiguously demonstrating the inherent fragility of the QSH phase
due to the activation of intervalley processes.} The dependence with
the adatom coverage is shown in panel Fig.~\ref{fig:03}(d), which
also shows the conductance at low energies is further degraded when
$t_{1}^{\prime}$ and $t_{3}^{\prime}$ are taken simultaneously\textcolor{black}{.}

\emph{Chern number in real-space}.\textemdash At this stage, we have
firmly established that edge states in adatom-decorated graphene are\emph{
intrinsically unstable} due to\textcolor{black}{\emph{ hitherto neglected}}\textcolor{black}{{}
}\textcolor{black}{\emph{hoppings}}\textcolor{black}{, $t_{1}^{\prime}$
and $t_{3}^{\prime}$ }(Fig.~\ref{fig:02}\textcolor{black}{). As
shown by our multi-scale theory bridging advanced first-principles
calculations and accurate TB models, such hoppings are typically one
order of magnitude larger than the chiral term inducing the topological
phase ($t_{2}^{\prime\prime}$)}.\textcolor{black}{{} To investigate
the onset of the topological phase transition in more detail, we evaluate
the spin Chern number of bulk states, defined by $\mathcal{C}_{s}=C_{\uparrow}-C_{\downarrow}$,
where $C_{\uparrow}=-C_{\downarrow}\equiv\mathcal{C}$ is the first
Chern integer \cite{QSH_Sheng_06}. To compute $\mathcal{C}$ for
disordered configurations of adatoms, we employ an efficient gauge-invariant
approach developed in Refs.~\cite{Chern_Fukui_05,Chern_Zhang_13}.
This allow us to assess the topological order for realistic TB parameters
(i.e., no $t_{n}$ rescaling is required). In Fig.}~\ref{fig:03}(e),
we show the dependence of the average Chern number $\bar{\mathcal{C}}$
on the adatom coverage. The robustness of $\bar{\mathcal{C}}$ with
respect to $t_{1}^{\prime\prime}$ and its quick suppression at low
concentrations when \textcolor{black}{$t_{1}^{\prime}$ and $t_{3}^{\prime}$}
are turned on is in perfect agreement with the previous conclusions
based on quantum transport simulations in nanoribbon geometry. When
all adatom-induced interactions are considered on equal footing, the
onset to the transition to a topologically trivial phase is found
to occur around 5\% for typical values of the parameters. When approaching
1-2\% coverage, the fluctuations are increasingly larger predicting
the closing of the topological gap in the relevant experimental regime.
The strong dependence of $\bar{\mathcal{C}}$ on the adatom coverage
is reminiscent of Anderson topological insulators \cite{Li_TAInsulators_09},
for which the character of the insulating ground state is known to
critically depend on the disorder strength. 

\textcolor{black}{Our findings have several important consequences.
The absence of topological gap signatures on recent measurements in
adatom-decorated graphene \cite{QSH_Exp_Adatom_1,QSH_Exp_Adatom_2,QSH_Exp_Adatom_3,QSH_Exp_Adatom_4,QSH_Exp_Adatom_5}
is naturally explained by valley mixing processes beyond simple model
Hamiltonians. The adatom-induced intervalley scattering uncovered
in this work can be mitigated by increasing the spatial range of the
interactions mediated by the adsorbate, thereby providing a possible
path towards the engineering of a quantum spin Hall insulator in graphene,
}\textcolor{black}{\emph{i.e.}}\textcolor{black}{, its decoration
with dilute heavy nano-particles. These results highlight the importance
of seamless multi-scale approaches bridging first-principles parameterized
model Hamiltonians and large-scale quantum transport calculations
for the predictive modelling of adatom\textendash host systems. }

Data availability statement.\textemdash The data underlying this paper
are available from the Figshare database \cite{figshare}.

\emph{Acknowledgments.}\textemdash We acknowledge support the\textbf{
}Engineering and Physical Sciences Research Council (EPRSC) under
Grants No. EP/N005244/1 (J.L.), and EP/K003151/1 and EP/P023843/1
(K.M.), the Thomas Young Centre under Grant No. TYC-101 (J.L.), the
Brazilian funding agency CAPES under Project No. 13703/13-7 (F.J.S.),
and the European Research Council (ERC) under the ERC-consolidator
Grant No. 681405-DYNASORE (F.J.S.). Via J.L. and K.M.'s membership
of the UK's HEC Materials Chemistry Consortium, which is funded by
EPSRC (EP/L000202), this work used the ARCHER UK National Supercomputing
Service. DAB acknowledges the support from Mackpesquisa and FAPESP
under grant 2012/50259-8. E.V.C. acknowledges partial support from
FCT-Portugal through Grant No. UID/CTM/04540/2013. A.F. gratefully
acknowledges the support from the Royal Society through a Royal Society
University Research Fellowship and partial funding from EPSRC (Grant
No. EP/N004817/1). F.J.S thanks Filipe S. M. Guimarães for the support
and helpful discussions. E.V.C. and A.F. thank M.P. López-Sancho and
M.A.H. Vozmediano for useful discussions.

\begin{widetext}

\section{First-Principles Calculations\label{sec:First-Principles-Calculations}}

\subsection{Framework}

We carry out \emph{ab initio} density-functional theory (DFT) calculations
for thallium (Tl) atoms adsorbed on graphene. We employ a $4\times4$
graphene supercell containing 32 carbon (C) atoms and a single Tl
adatom located above a hollow site, i.e. above the center of a graphene
hexagon. Our calculations are carried out within the planewave-pseudopotential
approach as implemented in the QUANTUM ESPRESSO~\cite{supp1_QuantumEspresso}
software package. For the electron-ion interaction, we employ fully-relativistic,
multiple-projector, normconserving pseudopotentials, that were generated
with the ONCVPSP code~\cite{supp2_hamann2013optimized}. For the
Tl pseudopotential, the $5s$ and $5p$ semicore states were included
in the valence states. We further use the the PBE exchange-correlation
energy functional~\cite{supp3_perdew1996generalized}, a $4\times4\times1$
Monkhorst-Pack grid to sample the Brillouin zone of the supercell~\cite{supp4_monkhorst1976special},
a 60~Ry plane-wave cutoff, a Gaussian smearing of 0.01~Ry and a
separation of 20~Å between periodic images of the graphene sheet,
which we assume to be stacked in the z-direction.

\subsection{Fully-relativistic structure relaxation }

We first relax the structure to determine the atomic ground state
configuration. Interestingly, we observe significant differences for
the height of the Tl atom above the graphene sheet for scalar-relativistic
and fully-relativistic DFT calculations: while the scalar-relativistic
approach yields at Tl height of 2.58~Å above the graphene plane (defined
as the average z-coordinate of all C atoms), the fully-relativistic
approach predicts a significantly larger height of 2.87~Å. Importantly,
this increase of the Tl heights results in a \emph{reduction} of the
band gap at the Dirac point from 24~meV to 13~meV. 
\begin{table}[b]
\begin{centering}
\begin{tabular}{|c|c|}
\hline 
method  & IP\tabularnewline
\hline 
\hline 
SR-KS  & 8.19\tabularnewline
\hline 
SR-GW  & 5.31\tabularnewline
\hline 
SR-$\Delta$SCF  & 5.17\tabularnewline
\hline 
FR-KS  & 9.28\tabularnewline
\hline 
FR-GW  & 5.95\tabularnewline
\hline 
FR-$\Delta$SCF  & 5.95\tabularnewline
\hline 
Experiment  & 6.1\tabularnewline
\hline 
\end{tabular}
\par\end{centering}
\caption{\label{table:tl_atom} Ionization potential of a thallium atom from
different methods. Here, SR-KS and FR-KS denote calculations, where
the ionization potential is estimated from the scalar-relativistic
and fully-relativistic DFT Kohn-Sham energies, respectively. SR-$\Delta$SCF
and FR-$\Delta$SCF denote IP estimates obtained from the total energy
difference of scalar-relativistic and fully-relativistic DFT calculations
for the neutral atom and the ion. SR-GW and FR-GW denote the ionization
potential from scalar-relativistic and fully-relativistic GW calculations.
Experimental data from Ref.~\cite{supp5_CRC}. All energies are given
in eV.}
\end{table}

\subsection{GW corrections}

To overcome the well-known band gap problem of DFT~\cite{supp6_LouieHybertsen}
and accurately determine the electronic structure of the adatom-graphene
system, we carry out a fully-relativistic GW calculation using the
BerkeleyGW program package~\cite{supp7_BGWpaper}. As a test, we
first carry out calculations on an isolated Tl atom and an unperturbed
graphene sheet. We place the Tl atom in a cubic unit cell with a linear
extent of 20~bohr. The dielectric matrix is calculated using a 15~Ry
plane-wave cutoff and 700 bands and then extended to finite frequencies
using a generalized plasmon-pole model~\cite{supp6_LouieHybertsen}.
Moreover, we employ a spherically truncated Coulomb interaction to
avoid spurious interactions between periodic images and a static remainder
correction to include contributions from high-energy unoccupied states
in the self-energy~\cite{supp8_StaticRemainder}. To obtain the ionization
potential (IP) of the neutral atom, we consider a Tl ion with one
less electron than the neutral atom~\footnote{Carrying out GW calculations for a neutral Tl atom is conceptually
difficult because it has a partially filled 6p shell. The negatively-charged
ion, on the other hand, has a closed-shell configuration and calculations
are straightforward.} and compute its electron affinity. Table~\ref{table:tl_atom} summarizes
our results and shows that the IP from fully-relativistic GW \emph{agrees
well with the experimental data}, while the estimate obtained from
the Kohn-Sham orbital energy of fully-relativistic DFT deviates by
more than 3~eV from experiment. Note also that fully-relativistic
GW predicts a significantly smaller value of 0.95~eV for the spin-orbit
splitting of the $6p$ states compared to fully-relativistic DFT,
which predicts 1.43~eV. We also computed the GW correction to the
DFT IP of unperturbed graphene and find it to be only 0.1~eV. This
analysis clearly shows that a GW calculation is necessary to reliably
describe the level alignment of the Tl-graphene system as the accuracy
of DFT-PBE for the IP of localized systems, such as atoms and molecules,
is significantly worse than for extended systems, such as graphene.

Finally, we carry out a fully-relativistic GW calculation for the
combined Tl-graphene system. We use a plane-wave cutoff of 15 Ry for
the dielectric matrix, a sum over 1000 states, a slab-truncated Coulomb
interaction, a static remainder correction, a generalized plasmon-pole
model and a $6\times6\times1$ kpoint grid. Fig.~\ref{fig:bands}(a)
shows the \emph{ab initio} GW band structure (red dots) and compares
it to the \emph{ab initio} DFT result. In the graph, the bands below
the Dirac point are graphene states. The flat bands above the Dirac
point, however, derive from Tl p-states. We observe that the GW correction
reduces the separation between the Dirac point and the lowest Tl $p$-state
by several 100 meV. Most importantly, we find that GW increases the
band gap at the Dirac point from its DFT values of 13 meV to 19 meV.

\begin{figure}
\centering{}\includegraphics[width=10cm]{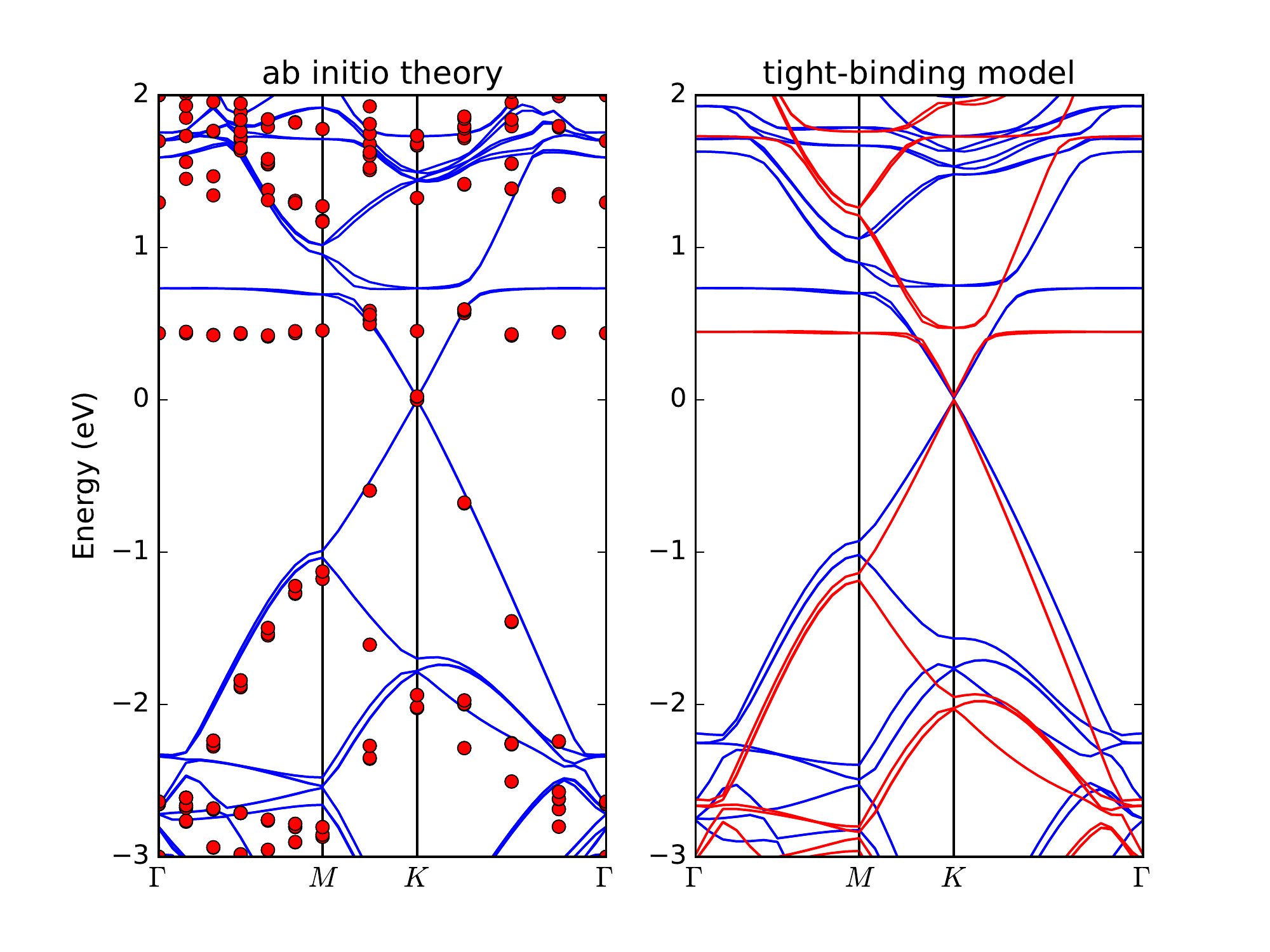} \caption{Fully-relativistic band structure of graphene with a thallium adatom.
Left: \emph{ab initio} DFT (blue curve) and GW (red dots) results.
Right: tight-binding results with DFT parameters (blue curve) and
GW parameters (red curve).}
\label{fig:bands} 
\end{figure}

\section{Tight-Binding Models\label{sec:Tight-Binding-Models}}

\subsection{Wannierization\label{subsec:Wannierization}}

We employ the WANNIER90 code to construct maximally localized Wannier
functions (MLWFs) for Tl-decorated graphene \cite{supp9_W_Mostofi_08,supp10_W_Marzari_97,supp11_W_Souza_01}.
The initial band-structure calculations are performed as described
below in QUANTUM ESPRESSO \cite{supp1_QuantumEspresso}. This approach
has been applied previously to pristine graphene \cite{supp12_W_Jung_13}.
Here we extend the approach to non-collinear calculations on Tl-decorated
graphene. A fully self-consistent calculation is performed on the
$4\times4$ supercell of graphene with a single Tl adatom using a
$4\times4\times1$ Monkhorst-Pack grid to sample the Brillouin zone.
With the self-consistent charge density held fixed we perform non-self
consistent calculations using a $3\times3\times1$ Monkhorst-Pack
grid on 986 bands. The overlap matrices and projections are calculated
using the PW2WANNIER90 routine of QUANTUM ESPRESSO. For the MLWF calculation
we consider atom-centred $p_{z}$ functions on all C sites and $p_{z}$,
$p_{y}$ and $p_{x}$ functions on Tl (i.e. 70 Wannier functions in
total accounting for spin). WANNIER90 is used to determine the MLWFs
with the following parameters for disentanglement: frozen inner window:
-4.0 to 2.0~eV, outer window -10.50 to 14.00~eV and maximum number
of iterative steps: 2000. Typical Wannier function spreads for the
optimised orbitals are $<1$~Å$^{2}$ (C) and $<5$~Å$^{2}$(Tl).

\begin{figure}
\begin{centering}
\includegraphics[width=0.3\textwidth]{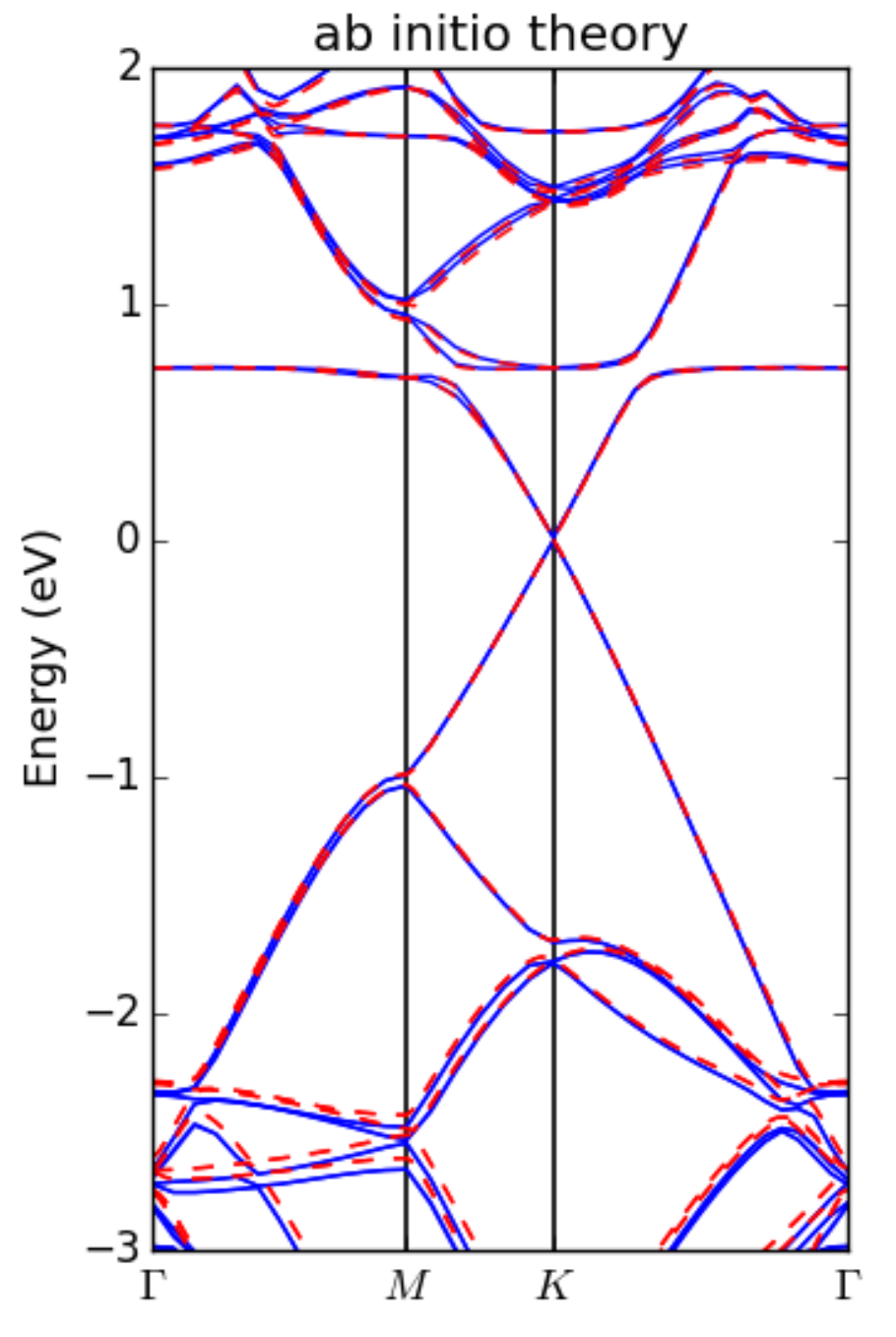} 
\par\end{centering}
\caption{\label{fig:MLWFs}DFT band structure of Tl-decorated graphene: plane-wave
(solid blue line) versus MLWFs (dashed red line). A vertical shift
is applied so that the Dirac point is positioned at zero energy. }
\end{figure}

As shown in Fig.~\ref{fig:MLWFs} the MLWFs obtained by the procedure
outlined above describe the band structure in very good agreement
with the plane wave basis especially near the region of interest (i.e.,
close to the Dirac point). Hamiltonian matrix elements between the
atom-centred MWLFs are computed and used to obtain an initial guess
for the parameters in our minimal model. The first, second and third
nearest neighbor C-C hopping parameters are computed for a pristine
graphene supercell (without the Tl adatom) as appropriate for the
dilute limit. Table I summarises the parameters extracted from the
Wannier fit to the DFT calculation as well as those obtained by fitting
directly to the DFT minimal model (see next section). $-\delta\mu$
(the local shift of the $p_{z}$-orbital energies on C atoms neighboring
Tl relative to pristine graphene) is estimated by taking the difference
between the site energies of first and third nearest C atoms to the
Tl adatom. In addition, our Wannierization calculations show that
spin\textendash dependent hopping integrals between Tl and the adjacent
C atoms are negligible, justifying the neglect of those terms in the
minimal model (see next section). 
\begin{center}
\begin{table}
\begin{centering}
\begin{tabular}{|c|c|c|c|c|c|c|c|c|c|c|}
\cline{2-11} 
\multicolumn{1}{c|}{} & $\tau_{0}$  & \textbf{$\tau_{1}$}  & \textbf{$\lambda$}  & \textbf{$\lambda^{\prime}$}  & \textbf{$\epsilon_{0}$}  & \textbf{$\epsilon_{1}$}  & \textbf{$t$}  & \textbf{$t^{\prime}$}  & \textbf{$t^{\prime\prime}$}  & \textbf{$\delta\mu$}\tabularnewline
\hline 
DFT Wannier  & ~~~-1.07~~~  & ~~~-0.59~~~  & ~~~0.37~~~  & ~~~0.19~~~  & ~~~1.65~~~  & ~~~1.28~~~  & ~~~2.87~~~  & ~~~-0.24~~~  & ~~~0.26~~~  & ~~~0.26~~\tabularnewline
\hline 
\end{tabular}
\par\end{centering}
\caption{Parameters (in eV) extracted from the MLWF fit to the DFT calculation.}
\end{table}
\par\end{center}

\subsection{Minimal graphene\textendash adatom model\label{subsec:Minimal_TB}}

Following Weeks \emph{et al.}~\cite{QSH_G_Adatom_Weeks_11}, we parameterize
a tight-binding Hamiltonian informed by the first-principles calculations.
The Hamiltonian is separated into a graphene contribution ($H_{g}$)
and a local adatom ($H_{a}$) and hybridization ($V_{ga}$) terms.
The accurate description of $\pi$ bands ($H_{g}$) requires a third-nearest
neighbor approximation \cite{supp12_W_Jung_13}. For the graphene\textendash adatom
interaction it suffices considering the nearest neighbors {[}Sec.\,(\ref{subsec:GrapheneOnlyTB}){]}.
Time-reversal symmetry and invariance with respect to the point group
$C_{6v}$ constrain the Hamiltonian to the following form\footnote{Symmetry also allows spin-dependent tunneling processes, e.g., $\alpha_{m}\Omega_{m}^{\dagger}s_{z}d_{m}$
and $\beta_{m}\Omega_{m}^{\dagger}s_{+}d_{m-1}$. However, these corresponding
hopping integrals are extremely small (Sec.~\ref{subsec:Wannierization}),
and thus such terms can be safely omitted.}: 
\begin{align}
H_{g} & =-t\sum_{\langle ij\rangle}\left(c_{\boldsymbol{r}_{i}}^{\dagger}c_{\boldsymbol{r}_{j}}+\textrm{H.c.}\right)+t^{\prime}\sum_{\langle\langle ij\rangle\rangle}\left(c_{\boldsymbol{r}_{i}}^{\dagger}c_{\boldsymbol{r}_{j}}+\textrm{H.c.}\right)+t^{\prime}\sum_{\langle\langle\langle ij\rangle\rangle\rangle}\left(c_{\boldsymbol{r}_{i}}^{\dagger}c_{\boldsymbol{r}_{j}}+\textrm{H.c.}\right)-\delta\mu\sum_{j=1}^{6}c_{\bm{r}_{j}}^{\dagger}c_{\bm{r}_{j}},\label{eq:Hg}\\
H_{a} & =\sum_{m=0,\pm1}\epsilon_{|m|}d_{m}^{\dagger}d_{m}+\lambda\left(d_{1}^{\dagger}s^{z}d_{1}-d_{-1}^{\dagger}s^{z}d_{-1}\right)+\sqrt{2}\lambda^{\prime}\left(d_{0}^{\dagger}s^{-}d_{-1}+d_{0}^{\dagger}s^{+}d_{1}+\textrm{H.c.}\right),\label{eq:Ha}\\
V_{ga} & =-\sum_{m=0,\pm1}\,i^{m}\tau_{|m|}\Omega_{m}^{\dagger}d_{m}+\textrm{H.c.}\,.\label{eq:Hc}
\end{align}
There are $10$ parameters in (\ref{eq:Hc}): the graphene hoppings
$t$, $t'$ and $t''$, the chemical potential change on C atoms next
to the Tl $\delta\mu$, the Tl p-state energies $\epsilon_{0}$ and
$\epsilon_{\pm1}$, the Tl spin-orbit parameters $\lambda,\lambda^{\prime}$
(we assume $\lambda^{\prime}=\lambda$) and the C-Tl hoppings $t_{0}$
and $t_{\pm1}$. We adjust these parameters until the tight-binding
band structures reproduce the \emph{ab initio} DFT and GW results.
Our final parameter values are given in Table~\ref{table:tb} and
the corresponding band structures are shown in Fig.~\ref{fig:bands}(b).

\begin{table*}[ht]
\begin{centering}
\begin{tabular}{c|c|c|c}
\multicolumn{1}{c}{} & \multicolumn{1}{c}{} & \multicolumn{1}{c|}{} & \tabularnewline
~~~~~~~~~~~  & ~~~DFT minimal~~~  & DFT Wannier  & ~~~GW minimal~~~\tabularnewline
\hline 
$\delta\mu$  & 0.83  & 0.26  & 0.28\tabularnewline
$t$  & 2.95  & 2.87  & 3.36\tabularnewline
$t^{\prime}$  & -0.12  & -0.24  & 0.10\tabularnewline
$t^{\prime\prime}$  & -0.25  & 0.26  & -0.19\tabularnewline
$\epsilon_{0}$  & \textbf{1.59}  & \textbf{1.65}  & \textbf{1.30}\tabularnewline
$\epsilon_{\pm1}$  & \textbf{1.45}  & \textbf{1.28}  & \textbf{0.82}\tabularnewline
$\tau_{0}$  & \textbf{-1.27}  & \textbf{-1.07}  & \textbf{-2.36}\tabularnewline
$\tau_{\pm1}$  & \textbf{-0.59}  & \textbf{-0.59}  & \textbf{-0.44}\tabularnewline
$\lambda$  & \textbf{0.33}  & \textbf{0.37}  & \textbf{0.51}\tabularnewline
\hline 
\end{tabular}
\par\end{centering}
\caption{\label{tab:Parameters}Minimal tight-binding models versus DFT Wannier
for graphene with a single thallium adatom. DFT and GW minimal denote
the parameters (in eV) determined by fitting to fully-relativistic
\emph{ab initio} DFT and GW band structures. }

\label{table:tb} 
\end{table*}

The DFT optimized parameters are in excellent agreement with the hopping
integrals obtained through the Wannierization procedure (Table~\ref{table:tb}).
This is remarkable given that the Wannierization scheme considers
\emph{all} hopping integrals between any pair of orbitals, whereas
the minimal model {[}Eq.~(\ref{eq:Hg})-(\ref{eq:Hc}){]} is truncated.
The discrepancy in the values of $\delta\mu$ is expected given that
the minimal model restricts the on-site potential to the six C atoms
adjacent to Tl, which then overestimates the depth of the potential
well.

\subsection{Decimation of the adatom degrees of freedom \label{subsec:GrapheneOnlyTB}}

We use standard degenerate perturbation theory to derive an accurate\emph{
graphene-only} Hamiltonian parameterized by our first-principles calculations.
The effective interaction induced by a \emph{single} adatom on graphene
is 
\begin{equation}
\Sigma_{P}=\sum_{m,s}\sum_{m',s'}\left[\mathcal{V}^{\dagger}\,(\epsilon-\mathcal{H}_{\textrm{ad}})^{-1}\mathcal{V}\right]_{(m,s),(m',s')}\,\Omega_{m,s}^{\dagger}\thinspace\Omega_{m',s'}\,,\label{eq:H_imp_general}
\end{equation}
with $\hat{\mathcal{H}}_{\textrm{ad}\,(m,s),(m^{\prime},s^{\prime})}=_{\mathcal{A}}\langle m,s|H_{a}|m',s'\rangle_{\mathcal{A}}$,
$\hat{\mathcal{V}}_{(m,s),(m^{\prime},s^{\prime})}=_{\mathcal{A}}\langle m,s|V_{ga}|m',s'\rangle_{\mathcal{P}}$
and where $\mathcal{A}$ ($\mathcal{P}$) spans the adatom (graphene)
subspace of states with definite angular momentum \cite{SHE_G_Pachoud_14-1}.
From Eqs.~(\ref{eq:Ha})-(\ref{eq:Hc}) we find: 
\begin{align}
\mathcal{H}_{\textrm{ad}} & =\left(\begin{array}{cccccc}
\epsilon_{1}+\lambda & 0 & 0 & 0 & 0 & 0\\
0 & \epsilon_{1}-\lambda & \sqrt{2}\lambda & 0 & 0 & 0\\
0 & \sqrt{2}\lambda & \epsilon_{0} & 0 & 0 & 0\\
0 & 0 & 0 & \epsilon_{0} & \begin{array}{c}
\sqrt{2}\lambda\end{array} & 0\\
0 & 0 & 0 & \sqrt{2}\lambda & \epsilon_{1}-\lambda & 0\\
0 & 0 & 0 & 0 & 0 & \epsilon_{1}+\lambda
\end{array}\right),\quad\mathcal{V}=-\left(\begin{array}{cccccc}
i\tau_{1} & 0 & 0 & 0 & 0 & 0\\
0 & i\tau_{1} & 0 & 0 & 0 & 0\\
0 & 0 & \tau_{0} & 0 & 0 & 0\\
0 & 0 & 0 & \tau_{0} & 0 & 0\\
0 & 0 & 0 & 0 & i\tau_{1} & 0\\
0 & 0 & 0 & 0 & 0 & i\tau_{1}
\end{array}\right)\,,\label{eq:H_V_explicit}
\end{align}
where the states have been arranged in ascending total angular momentum,
i.e., $\{|-1,\downarrow\rangle_{\mathcal{A}},\,...\,,|1,\uparrow\rangle_{\textrm{\ensuremath{\mathcal{A}}}}$
and $\{|-1,\downarrow\rangle_{\mathcal{P}},\,..,|1,\downarrow\rangle_{\mathcal{P}}\}$.
Evaluation of the effective Hamiltonian Eq.~(\ref{eq:H_imp_general})
results in: 
\begin{eqnarray}
\Sigma_{P} & =6\times\left(\begin{array}{cccccc}
X_{-1} & 0 & 0 & 0 & 0 & 0\\
0 & X_{1} & -i\lambda_{R} & 0 & 0 & 0\\
0 & i\lambda_{R} & X_{0} & 0 & 0 & 0\\
0 & 0 & 0 & X_{0} & i\lambda_{R} & 0\\
0 & 0 & 0 & -i\lambda_{R} & X_{1} & 0\\
0 & 0 & 0 & 0 & 0 & X_{-1}
\end{array}\right) & \,,\label{eq:H_ad_explicit}
\end{eqnarray}
with 
\begin{equation}
\lambda_{R}=\frac{\sqrt{2}\lambda\tau_{0}\tau_{1}}{D_{\textrm{ad}}(\epsilon)},\,\quad X_{1}=\frac{\tau_{1}^{2}(\epsilon-\epsilon_{0})}{D_{\textrm{ad}}(\epsilon)},\,\quad X_{0}=\frac{\tau_{0}^{2}(\epsilon-\epsilon_{1}+\lambda)}{D_{\textrm{ad}}(\epsilon)},\,\quad X_{-1}=\frac{\tau_{1}^{2}}{6(\epsilon-\epsilon_{1}-\lambda)}\,,\label{eq:parameters}
\end{equation}
and 
\begin{eqnarray}
D_{\textrm{ad}}(\epsilon) & = & 6\left[\epsilon^{2}+\epsilon_{0}\left(\epsilon_{1}-\lambda\right)-\epsilon(\epsilon_{0}+\epsilon_{1}-\lambda)-2\lambda^{2}\right]\,.\label{eq:Den}
\end{eqnarray}
Changing Eq.~(\ref{eq:H_ad_explicit}) into the honeycomb site basis
$\{|\boldsymbol{r}_{j},s\rangle\}_{j\in P}$ yields the effective
graphene\textendash \emph{only} coupling Hamiltonian reported in the
main text (omiting on site terms): 
\begin{align}
\hat{\Sigma}_{P} & =\sum_{\langle ij\rangle\in P}c_{\boldsymbol{r}_{i}}^{\dagger}\left(t_{\text{1}}^{\prime}+\imath\zeta_{ij}t_{1}^{\prime\prime}s^{z}\right)c_{\boldsymbol{r}_{j}}+\sum_{\langle\langle ij\rangle\rangle\in P}c_{\boldsymbol{r}_{i}}^{\dagger}\left(t_{\text{2}}^{\prime}+\imath\zeta_{ij}t_{2}^{\prime\prime}s^{z}\right)c_{\boldsymbol{r}_{j}}+\sum_{\langle\langle\langle ij\rangle\rangle\rangle\in P}t_{\text{3}}^{\prime}c_{\boldsymbol{r}_{i}}^{\dagger}c_{\boldsymbol{r}_{j}}+\lambda_{R}\sum_{i,j\in P}c_{\bm{r}_{i}}^{\dagger}\Lambda_{ij}c_{\bm{r}_{j}}\,,\label{eq:V_eff}
\end{align}
where 
\begin{equation}
t_{n}^{\prime}+\imath t_{n}^{\prime\prime}=e^{\imath n\pi/3}X_{1}+X_{0}+e^{-\imath n\pi/3}X_{-1}\,,\qquad\Lambda_{ij}=\Lambda_{ij}=\imath(\xi_{ij}s^{+}+\xi_{ij}^{*}s^{-})\,,\label{eq:def}
\end{equation}
The total effective tight-binding Hamiltonian for a graphene sheet
with an adatom at the $P$-th hollow site (restoring on-site terms)
reads 
\begin{equation}
H_{P}=H_{g}-(\delta\mu-t_{0})\sum_{i\in P}c_{\boldsymbol{r}_{i}}^{\dagger}c_{\boldsymbol{r}_{i}}+\Sigma_{P}\,.\label{eq:eff_Ham}
\end{equation}
The on-site energy can be trivially absorbed in the definition of
the local chemical potential $\delta\mu\rightarrow\delta\mu-t_{0}$
in Eq.~(\ref{eq:Hg}). Numerical studies showed that moderate values
of $\delta\mu$ do \emph{not} impact the topological phase \cite{QSH_G_Adatom_Weeks_11}.
The underlying reason is that uniform on-site energies lead to a \emph{scalar}
(intra-valley) potential, $u(\boldsymbol{r})=-3\,\delta\mu\,f_{P}(\boldsymbol{r})$
to leading order in $\mathbf{k}/K$, in the long wavelenght description
(see main text). The main effect of a moderate on site energy correction
$|\delta\mu|\ll6t$ is thus a trivial shift in the spectrum: $\delta\epsilon=-3n\delta\mu$,
where $n$ is the adatom coverage.

\section{Real-Space Quantum Transport Calculations\label{sec:Real-Space-Quantum-Transport}}

\subsection{Methodology}

\begin{figure}[ht]
\centering{}\includegraphics[width=0.5\textwidth]{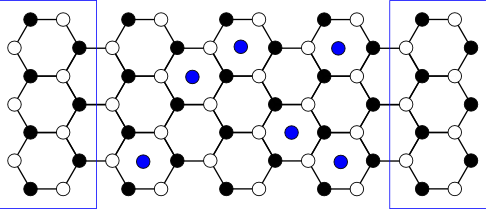}
\caption{Armchair graphene nanoribbon two-terminal device. Adatoms (blue) are
sprinkled over the central region.}
\label{fig:nanoribbon} 
\end{figure}

We consider an adatom-decorated graphene nanoribbon connected to pristine
armchair-graphene leads\footnote{The popular choice of armchair termination for studies of the QSH
phase in graphene is meant to eliminate the possibility of edge states
occuring already in the absence of SOC (albeit with a different origin).
Our armchair nanoribbons are made deliberately metallic ($W=2p+3$;
with $p\in\mathbb{N}$) and hence their conductance exibit a natural
plateau at low energies of width $\delta E=3t/W$ associated with
states delocalized through the nanoribbon. This confinement effect
puts a constraint on the smallest SOC gaps that can be probed by the
transport calculations: $\Delta_{\textrm{QSH}}\gtrsim\delta E$ (Refs.~\cite{supp13_GNR_Son_06,supp14_GNR_Zheng07}).} as depicted in Fig.~\ref{fig:nanoribbon}. To determine the quantum
transport characteristics, we employ the Landauer-Büttiker formalism~\cite{supp15_Buttiker_98,supp16_Landauer_70},
where the longitudinal conductance is expressed as \cite{supp17_Caroli_71,supp18_Datta_Book,supp19_Fisher_81}:
\begin{equation}
G=\frac{2e^{2}}{h}\textrm{Tr}\left[\left(\Sigma_{l}^{A}-\Sigma_{l}^{R}\right)\mathcal{G}_{lr}^{R}\left(\Sigma_{r}^{R}-\Sigma_{r}^{A}\right)\mathcal{G}_{lr}^{A}\right],\label{eq:landauer}
\end{equation}
where $\mathcal{G}_{lr}^{a}$ is Green's function connecting the left
and right interfaces of the central device and $\Sigma_{l(r)}^{a}$
is the self-energy of the left (right) contact in the retarded (R)\textemdash advanced
(A) sector. The central device with $N$ adatoms is modelled according
to the following Hamiltonian: 
\begin{equation}
H=-t\sum_{\langle ij\rangle}c_{\boldsymbol{r}_{i}}^{\dagger}c_{\boldsymbol{r}_{j}}+\sum_{P=1}^{N}\left[\sum_{\langle ij\rangle\in P}c_{\boldsymbol{r}_{i}}^{\dagger}\left(t_{\text{1}}^{\prime}+\imath\zeta_{ij}t_{1}^{\prime\prime}\hat{s}_{z}\right)c_{\boldsymbol{r}_{j}}+\sum_{\langle\langle ij\rangle\rangle\in P}c_{\boldsymbol{r}_{i}}^{\dagger}\left(t_{\text{2}}^{\prime}+\imath\zeta_{ij}t_{2}^{\prime\prime}\hat{s}_{z}\right)c_{\boldsymbol{r}_{j}}+\sum_{\langle\langle\langle ij\rangle\rangle\rangle\in P}t_{\text{3}}^{\prime}c_{\boldsymbol{r}_{i}}^{\dagger}c_{\boldsymbol{r}_{i}}\right]\,.\label{eq:H_quantum_transport}
\end{equation}
The calculation is carried out in two steps. Firstly, the pristine
semi-infinite armchair graphene-contacts are treated analytically.
Secondly, an efficient, highly-parallelizable recursive numerical
approach is employed to extract the Green's function of the central
device. The method consists of subdividing the central region in smaller
parts, whose Green's functions can be evaluated via direct inversion~\cite{supp20_Lewenkopf_13}.
The numerical evaluation of the Green's function for the central region,
together with the analytic treatment of the leads, allows to efficiently
compute the conductance of sizable graphene nanoribbons with several
millions of atoms.

\subsection{Bond current maps}

The bond current density $J_{ij}$ for up and down spins ($s=\pm1$)
is obtained from the variation rate of the charge density at a given
site, which is written in terms of the flux of electrons through all
of its bonds. For that, one needs to calculate the variation of the
expectation value of the number operator over time, leading to the
following expression for the current flowing from site $i$ to $j$
\cite{supp21_AntiPekka}: 
\begin{equation}
J_{ij}^{s}=\frac{4e}{\hbar}\,\Im\left[t_{ij}\mathcal{C}_{ij}^{s}\right],
\end{equation}
where $\mathcal{C}\equiv\imath G_{lr}^{R}(\Sigma_{r}^{R}-\Sigma_{r}^{A})G_{lr}^{A}$
is the correlation function and $t_{ij}$ is the hopping between sites
$i$ and $j$. The density maps borne out the crucial role played
by the new adatom-induced hoppings studied here (Fig.~\ref{fig:density-maps}).

\begin{figure}[H]
\begin{centering}
\includegraphics[width=0.5\columnwidth]{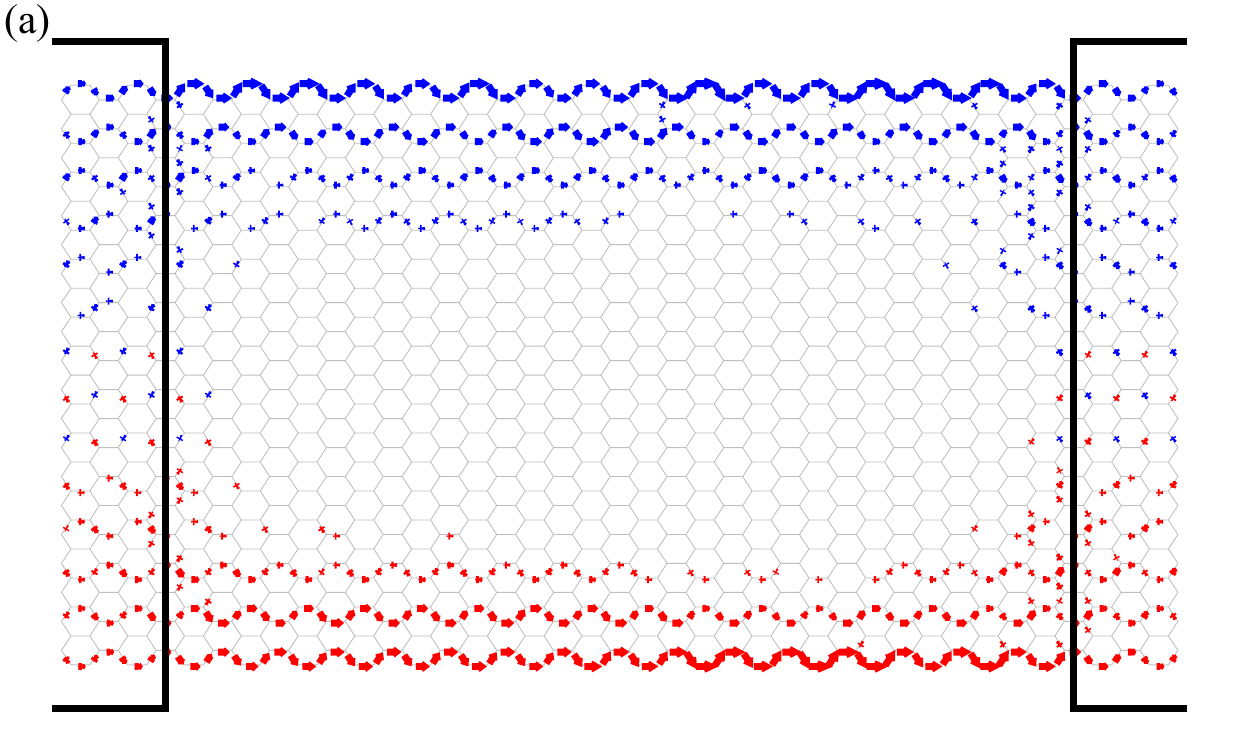} 
\par\end{centering}
\begin{centering}
\includegraphics[width=0.5\columnwidth]{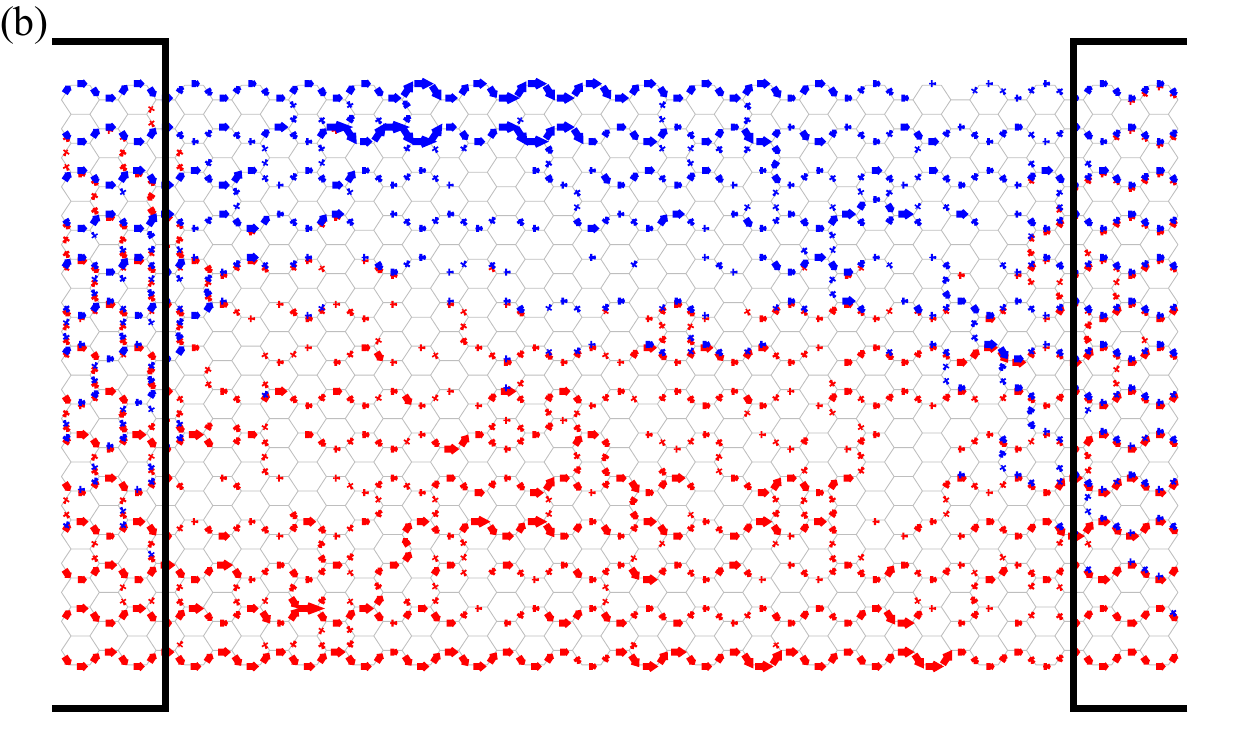} 
\par\end{centering}
\centering{}\includegraphics[width=0.5\columnwidth]{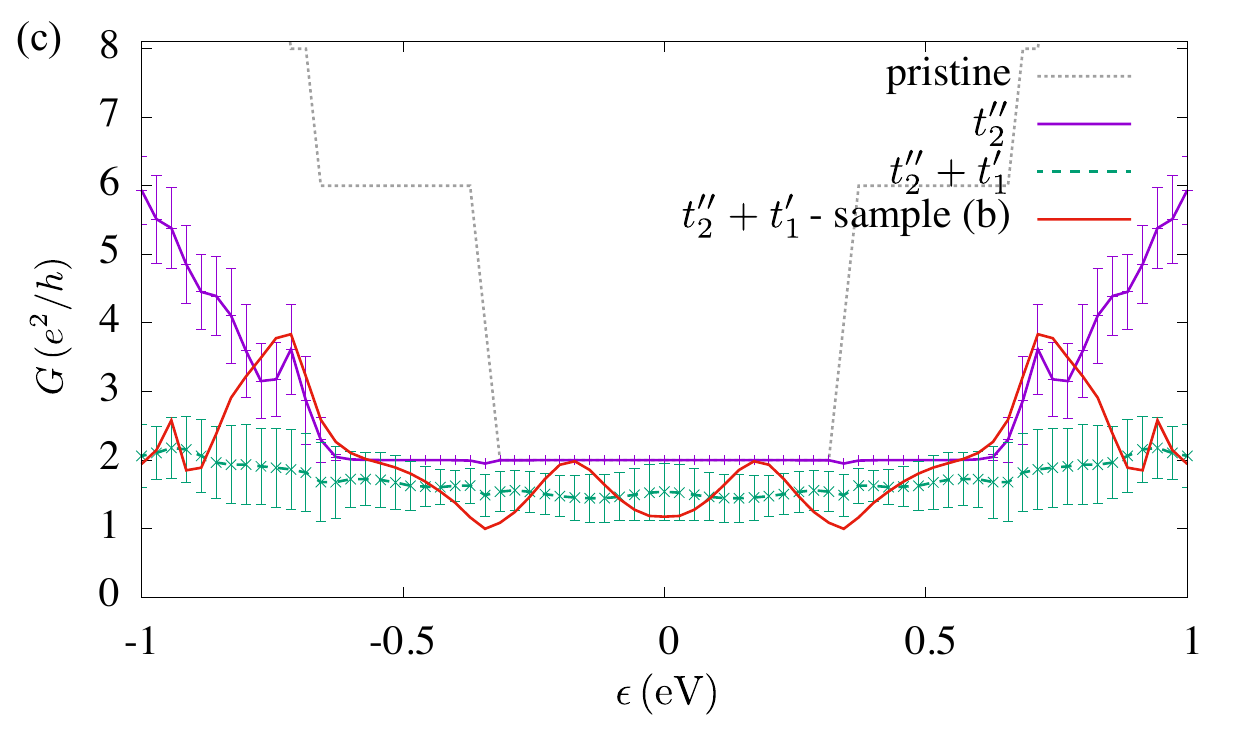}\caption{\label{fig:density-maps}Current density map $J_{ij}^{s}$ in a two-terminal
device with random adatoms {[}$s=\uparrow,\downarrow$ red (blue){]}.
(a) Helical edge states appear for artificial adatoms mediating only
a next-nearest neighbor spin\textendash orbit hopping, $t_{2}^{\prime\prime}$,
in graphene \cite{QSH_G_Adatom_Weeks_11}. (b) Helical edge states
are efficiently suppressed in realistic models incorporating the strong
nearest-neighbor hopping correction, $t_{1}^{\prime}$. (c) Energy
dependence of the conductance. Densities in a.u. Parameters (graphene):
$t=2.7$~eV and $t^{\prime}=t^{\prime\prime}=0$ ($W=L=20$). Parameters
(adatom): coverage of 50\% (24 samples), $t_{1}=-2.4$~eV, $t_{2}=-i0.23$~eV
and $t_{3}=0$.}
\end{figure}

\section{Real-Space Chern Number Calculations\label{sec:Real-Space-Chern}}

\subsection{Methodology}

We calculate the first Chern number $\mathcal{C}$ using a definition
which is valid for systems which are not translationaly invariant.
To that purpose we use twisted boundary conditions, 
\begin{equation}
u_{n}^{\bm{\theta}}(\mathbf{r}+N_{i}\mathbf{a}_{i})\equiv\left\langle \mathbf{r}+N_{i}\mathbf{a}_{i}\right.\left|u_{n}^{\bm{\theta}}\right\rangle =e^{i\theta_{i}}u_{n}^{\bm{\theta}}(\mathbf{r})\,,\label{eq:twisted-1}
\end{equation}
with $\bm{\theta}=(\theta_{1},\theta_{2})$ for the twist angles $0\leq\theta_{1},\theta_{2}\leq2\pi$,
and $H(\bm{\theta})\left|u_{n}^{\bm{\theta}}\right\rangle =E_{n}\left|u_{n}^{\bm{\theta}}\right\rangle $
for the system Halmiltonian $H(\bm{\theta})$ with twisted boundary
conditions $\bm{\theta}$, where $\mathbf{a}_{i}$ is a basis vector
of the direct lattice with $N_{i}$ lattice points in the direction
of $\mathbf{a}_{i}$. The ground state wave function $\left|\Psi^{\bm{\theta}}\right\rangle $
is obtained from single particle states $\left|u_{n}^{\bm{\theta}}\right\rangle $,
\begin{equation}
\left|\Psi^{\bm{\theta}}\right\rangle =\prod_{n=1}^{M}\varphi_{n}^{\dagger}(\bm{\theta})\left|0\right\rangle ,\:\textrm{where}\,\,\left|u_{n}^{\bm{\theta}}\right\rangle \equiv\varphi_{n}^{\dagger}(\bm{\theta})\left|0\right\rangle \,,\label{eq:twisted-2}
\end{equation}
with $E_{n}<E_{n+1}$, and we assume that only $M$ single particle
states are occupied.

If the ground state is non-degenerate and there is a finite energy
gap between the ground state energy and the excited states in the
bulk then the Chern number is well defined, 
\begin{equation}
\mathcal{C}=\frac{1}{2\pi}\int_{S_{\theta}}d\bm{\theta}\left[\nabla_{\bm{\theta}}\times\mathbf{A}(\bm{\theta})\right]_{z}\,,\label{eq:first_Chern}
\end{equation}
with 
\begin{equation}
\mathbf{A}(\bm{\theta})=\left\langle \Psi^{\bm{\theta}}\right|i\nabla_{\bm{\theta}}\left|\Psi^{\bm{\theta}}\right\rangle \,,\label{eq:Connection}
\end{equation}
where $S_{\theta}$ is the surface $0\leq\theta_{1},\theta_{2}\leq2\pi$.
To evaluate Eq.\,(\ref{eq:first_Chern}), we apply Fukui's method
\cite{Chern_Fukui_05} by discretizing the surface $S_{\theta}$ into
$L_{1}L_{2}$ points and then suming over the flux of individual plaquettes,
\begin{equation}
C=\frac{1}{2\pi}\sum_{l=1}^{L_{1}L_{2}}\arg\left(\left\langle \Psi^{\bm{\theta}_{l}}\right.\left|\Psi^{\bm{\theta}_{l}+\hat{1}}\right\rangle \left\langle \Psi^{\bm{\theta}_{l}+\hat{1}}\right.\left|\Psi^{\bm{\theta}_{l}+\hat{1}+\hat{2}}\right\rangle \left\langle \Psi^{\bm{\theta}_{l}+\hat{1}+\hat{2}}\right.\left|\Psi^{\bm{\theta}_{l}+\hat{2}}\right\rangle \left\langle \Psi^{\bm{\theta}_{l}+\hat{2}}\right.\left|\Psi^{\bm{\theta}_{l}}\right\rangle \right),\label{eq:ChernTheta-1}
\end{equation}
where $\hat{\mu}$ is a vector in the direction $\mu=1,2$ with magnitude
$|\hat{\mu}|=\sqrt{A(S_{\theta})}/N_{\mu}$, with $A(S_{\theta})$
the area of the surface $S_{\theta}$. The advantage of this approach
is that $C$ is strictly an integer for arbitrary lattice spacing,
and it should rapidly converge to $\mathcal{C}$ as one increases
the size of the lattice $L_{1}L_{2}$. We adopted the implementation
discussed in Ref.~\cite{Chern_Zhang_13}.

\subsection{Results}

In order to assess quantitatively the transition between topologically
trivial and non-trivial ground states, we have carried on a detailed
calculation of the topological index as a function of adatom coverage.
The results are shown in Fig.~\ref{fig:chern} from the lowest densities
we can simulate up to 10\% coverage. Since we calculate the Chern
number on a finite size lattice with $\mathcal{N}$ = $d\times d$
unit cells, the lowest density corresponds to a single adatoms and
$n$ takes the value $n=1/\mathcal{N}$.

\begin{figure}[ht]
\centering{}\includegraphics[width=0.7\textwidth]{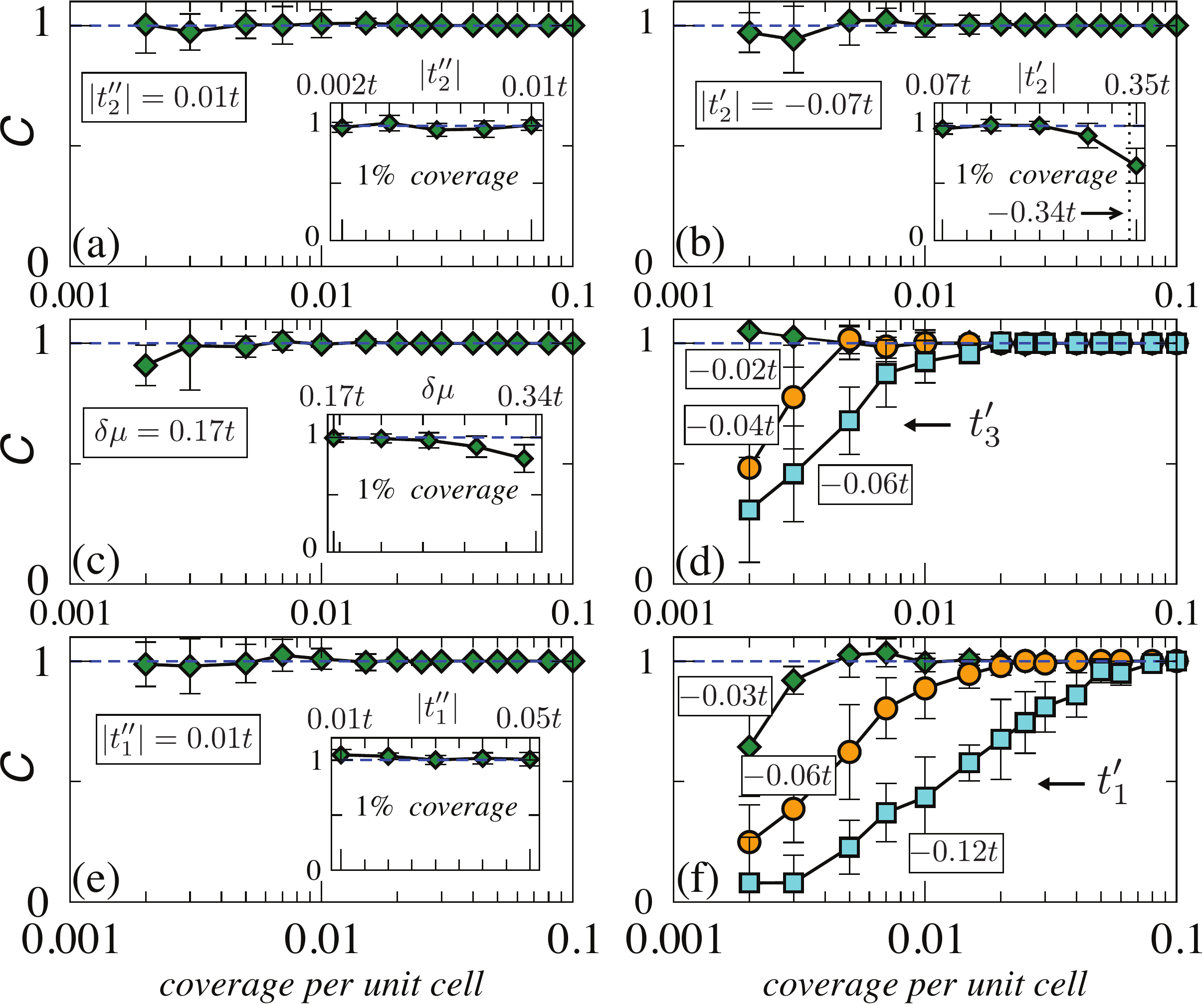}\caption{\label{fig:chern}Adatom coverage dependence of average Chern number:
(a) with only $t_{2}^{\prime\prime}\protect\neq0$ and (b)\textendash (f)
with $t_{2}^{\prime\prime}/t=-0.01$ and keeping the parameter indicated
in each panel fixed, while setting all others to zero. In the insets
we fix the coverage to 0.01 atoms per unit cell, and vary the parameter
indicated in each panel. }
\end{figure}

\end{widetext}

\begin{thebibliography}{10}
\bibitem{Note_Minimal_Model}Spin-dependent tunneling processes (e.g.,
$\propto\Omega_{m,\uparrow}^{\dagger}d_{m,\downarrow}$) have negligible
probability and thus are omitted. Moreover, crystal-field effects
are neglected \cite{QSH_G_Adatom_Weeks_11,SM,Cysne_18}.
\bibitem{Comment}The low-energy theory for generic physisorbed adatoms
was first derived in Ref.\,\cite{SHE_G_Pachoud_14}.
\end{thebibliography}
\end{document}